\documentclass[times,tighten,twocolumn]{aastex6}
\usepackage{graphicx}

\newcommand{\oiii}{[O {\sc iii}]}
\newcommand{\fx}{\ifmmode {f_{\rm 2-10}} \else $f_{\rm 2-10}$ \fi}
\newcommand{\Lx}{\ifmmode {L_{\rm 2-10}} \else $L_{\rm 2-10}$ \fi}
\newcommand{\rblr}{\ifmmode {R_{\rm BLR}} \else $R_{\rm BLR}$ \fi}
\newcommand{\foiii}{\ifmmode {f_{\rm [OIII]}} \else $f_{\rm [OIII]}$ \fi}
\newcommand{\fmeanfwhm}{\ifmmode {f_{\rm FWHM}{\rm(mean)}} \else $f_{\rm FWHM}{\rm(mean)}$ \fi}
\newcommand{\frmsfwhm}{\ifmmode {f_{\rm FWHM}{\rm(rms)}} \else $f_{\rm FWHM}{\rm(rms)}$ \fi}
\newcommand{\fmeansigma}{\ifmmode {f_{\rm \sigma}{\rm(mean)}} \else $f_{\rm \sigma}{\rm(mean)}$ \fi}
\newcommand{\frmssigma}{\ifmmode {f_{\rm \sigma}{\rm(rms)}} \else $f_{\rm \sigma}{\rm(rms)}$ \fi}
\newcommand{\hb}{\ifmmode {\rm H\beta} \else H$\beta$ \fi}
\newcommand{\ha}{\ifmmode {\rm H\alpha} \else H$\alpha$ \fi}
\newcommand{\msigma}{$M_{\bullet}$ -- $\sigma_{*}$}

\journalinfo{To appear in {\it The Astrophysical Journal Letters}.}

\begin{document}

\title{Hidden Broad-line Regions in Seyfert 2 Galaxies: from the spectropolarimetric perspective}

\author{Pu Du\altaffilmark{1}, 
Jian-Min Wang\altaffilmark{1, 2, 3}, and 
Zhi-Xiang Zhang\altaffilmark{1}}

\altaffiltext{1}{Key Laboratory for Particle Astrophysics, Institute of High Energy
Physics, Chinese Academy of Sciences, 19B Yuquan Road, Beijing 100049, China}
\email{dupu@ihep.ac.cn}

\altaffiltext{2}{National Astronomical Observatories of China, 
Chinese Academy of Sciences, 20A Datun Road, Beijing 100020, China}

\altaffiltext{3}{School of Astronomy and Space Science, University of Chinese Academy of Sciences, 19A Yuquan Road, Beijing 100049, China}

\begin{abstract}
The hidden broad-line regions (BLRs) in Seyfert 2 galaxies, which display broad emission 
lines (BELs) in their polarized spectra, are a key piece of evidence in support of the 
unified model for active galactic nuclei (AGNs). However, the detailed kinematics and 
geometry of hidden BLRs are still not fully understood. The virial factor obtained from 
reverberation mapping of type 1 AGNs may be a useful diagnostic of the nature of hidden 
BLRs in type 2 objects. In order to understand the hidden BLRs, we compile six type 2 
objects from the literature with polarized BELs and dynamical measurements of black hole 
masses. All of them contain pseudobulges. We estimate their virial factors, and find the 
average value is 0.60 and the standard deviation is 0.69, which agree well with the value 
of type 1 AGNs with pseudobulges. This study demonstrates that (1) the geometry and kinematics 
of BLR are similar in type 1 and type 2 AGNs of the same bulge type (pseudobulges), and 
(2) the small values of virial factors in Seyfert 2 galaxies suggest that, similar to type 1 
AGNs, BLRs tend to be very thick disks in type 2 objects.
\end{abstract}

  
\section{Introduction}

The unified model \citep{antonucci1993}, which originates from the detection of
broad emission lines (BELs) in the spectropolarimetric data of Seyfert 2 galaxy 
NGC 1068 \citep{antonucci1985}, proposes that the type 1 (showing both BELs 
and narrow emission lines) and type 2 (showing only narrow emission lines) 
active galactic nuclei (AGNs) are the same type of objects viewed from different 
angles. The hidden broad-line regions (BLRs) are observed in the polarized spectra 
of many type 2 AGNs (e.g., \citealt{antonucci1984, tran1992, young1996, moran2000, 
ramos2016}), and are thought to have scattering of BELs by the regions above the poles 
of accretion disks (called ``polar scatter regions") \citep{antonucci1985, 
antonucci1993}. However, the detailed geometry and kinematics of BLRs hidden in the 
centers of type 2 AGNs remain elusive and are not yet fully understood. 

The virial mass of black holes can be obtained by reverberation mapping (RM, e.g., 
\citealt{blandford1982, peterson1998, kaspi2000, bentz2009, du2014, du2015, du2016}). 
The RM technique provides a key component, the time lag between continuum and BEL 
response, that provides a sizescale due to the finite speed of light. There is a 
dilemma that we are not able to get the black hole mass without specifying the 
kinematics and structure of the BLR. Fortunately, we can obtain the black hole 
mass from galactic dynamics in type 2 AGNs, providing the opportunity to define 
the virial factor   
\begin{equation}  
\label{eqn:mass}  
f_{\rm vir}=\frac{M_{\bullet}}{M_{\rm vir}}
             =\frac{GM_{\bullet}}{V^2R_{\rm BLR}}
\end{equation}  
where $M_{\bullet}$ is the BH mass determined by galactic dynamics techniques, 
$M_{\rm vir}$ is the virial mass, $V$ is the width of BEL, $R_{\rm BLR}$ is the 
emissivity-weighted radius of BLR, $G$ is the gravitational constant. For type 1 
AGNs, $R_{\rm BLR}=c\tau$, where $c$ is the speed of light, $\tau$ is the time 
delay between the response of broad emission lines (mainly H$\beta$) to the variation 
of continuum which is normally measured from the peak of cross-correlation functions 
between the light curves of the continuum and emission lines. The value of virial 
factor is determined by the geometry and kinematics of BLR, and its evaluation leads 
to an indirect path to understand the nature of BLRs. Comparing statistically with 
the BH mass -- stellar velocity dispersion (\msigma) relation (e.g., 
\citealt{ferrarese2000, tremaine2002}), its average value is estimated as the order 
of unity for type 1 AGNs if the velocity $V$ is measured from the full-width-half-maximum 
(FWHM) of broad H$\beta$ lines (e.g., \citealt{onken2004, ho2014, woo2015}). However, 
a more recent work indicates that the virial factor in individual objects probably 
varies significantly, as demonstrated by reconstructing the geometry and kinematics 
of BLRs in a small sample using the Markov Chain Monte Carlo method (MCMC, 
\citealt{pancoast2014b}). This result means that the kinematics of BLRs or the viewing 
angles are diverse in these objects.

The polar scattering regions in type 2 AGNs provides an additional viewing angle, namely 
from the polar axes, similar to the angle of observation for type 1 AGN. Considering 
that the virial factor reflects the geometry and kinematics of BLRs, we can diagnose 
the hidden BLRs in type 2 AGNs by comparing the virial factors, obtained from their 
polarized spectra, to those of type 1 AGNs. Therefore, we search for type 2 AGNs with 
hidden BELs in polarized spectra and directly determined dynamical BH mass measurements, 
and estimate their $f_{\rm vir}$ factors.

\section{Virial Factors of Type 2 AGNs}
\label{sec:data}

We compile all of the type 2 AGNs which have both polarized BEL detections 
and direct dynamical BH mass measurements from the published literatures 
in order to check their virial factors. The objects which are identified 
definitely as low-ionization nuclear emission-line regions (LINER, e.g., 
\citealt{ho2008}) or those with uncertain BH mass measurements have been 
eliminated. There are a total of six sources in the final sample. The 
degree of polarization in type 2 AGNs is generally very low (less than 
several percent), in most cases, only broad \ha lines in the polarized 
spectra are visible (e.g., \citealt{antonucci1984, tran1992, young1996, 
moran2000, ramos2016}). We assume that broad \hb and \ha lines have the 
same widths in the polarized spectrum for each individual object, as 
has been demonstrated in several objects \citep{ramos2016}. Because the 
poor quality of polarized spectra would influence the measurements of 
the line dispersions (second momentum of the profile, $\sigma$), we only 
adopt the FWHM of polarized emission lines in the following analysis. 

\begin{figure}[!ht]
\includegraphics[width=0.47\textwidth]{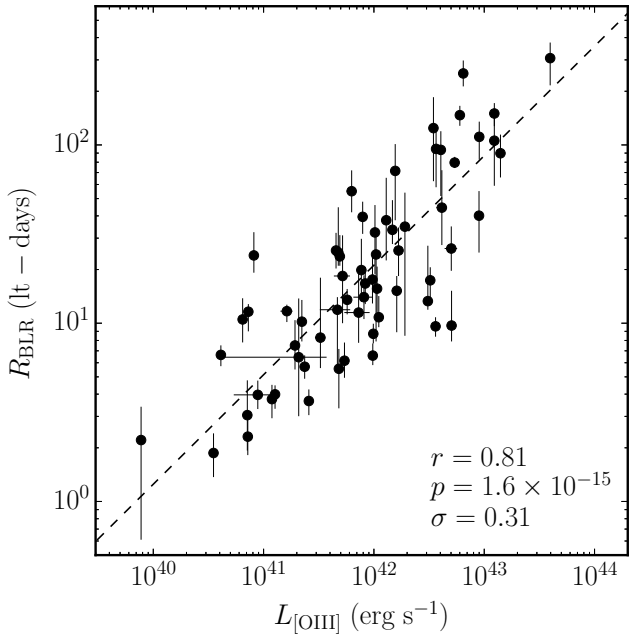}
\caption{Correlation between the luminosity of \oiii\ and $R_{\rm BLR}$. 
The points with error bars are the RM objects summarized in \cite{du2015, du2016}. 
The dotted line is the best fit. $r$ is the Pearson's correlation coefficient, 
$p$ is null-probability of no correlation and $\sigma$ is the standard deviation 
of residual after subtracting the best fit in logarithmic space. We adopt 
$\Lambda$CDM cosmology with $H_0=70~{\rm km~s^{-1}~Mpc^{-1}}$, 
$\Omega_{\Lambda}=0.7$ and $\Omega_{\rm M}=0.3$.}
\label{fig:loiii_tau}
\end{figure}

To estimate the radii of BLRs in Equation \ref{eqn:mass}, instead of the RM method 
as employed for type 1 AGNs, we use of two approaches: (1) the empirical relation 
between the time delay and the luminosity of \oiii$\lambda$5007 and (2) the empirical 
relation with the hard X-ray luminosity in 2-10 keV found in the RM samples. For the 
first approach, we compile the luminosities of \oiii\ and the time delays of \hb for 
all of the type 1 AGNs that have RM observations summarized in \cite{du2015, du2016}. 
It should be noted that the narrow-line regions (NLRs) in type 1 AGNs suffer 
local extinction of E(B-V) $\sim0.2-0.3$ (e.g., \citealt{netzer2006, vaona2012}) 
in addition to the Galactic extinction \citep{schlafly2011}. The line extinction 
should be corrected for the RM objects. Considering that the Balmer decrements 
of narrow lines for some RM objects are difficult to determine accurately because 
of their Lorentzian-like \ha and \hb profiles (encompassing both narrow and 
broad components)\footnote{The Lorentzian profiles of the entire \ha and \hb 
emission lines and the too weak narrow lines for some PG quasars in \cite{kaspi2000} and 
most of the AGNs with high accretion rates in \cite{du2014,du2015,du2016} make it 
difficult to de-blend the narrow components from the broad components very reliably, 
and additionally, difficult to measure the reddening in their narrow-line regions using 
Balmer decrements.}, we adopt the mean E(B-V) of 0.28, which is obtained from the Data 
Release 7 quasar sample of Sloan Digital Sky Survey (SDSS, \citealt{shen2011}), to 
correct the intrinsic extinction in those RM objects. Using the FITEXY algorithm 
\citep{press1992} for linear regression, the resulting correlation (shown in Figure 
\ref{fig:loiii_tau}) is 
\begin{equation}
\label{eqn:loiii_tau}
\log R_{\rm BLR} = -(24.475\pm0.684) + (0.614\pm0.016) \log L_{\rm [OIII]},
\end{equation}
where $L_{\rm [OIII]}$ is the extinction-corrected luminosity of \oiii. Equation 
\ref{eqn:loiii_tau} can be used to estimate the BLR radius. In order to avoid 
the potential contamination of star formation activities in \oiii\ fluxes, which 
would result in uncertainties in \rblr estimates to some extent, we also adopt 
hard X-ray luminosity in 2-10 keV (\Lx, the relation of \Lx versus \rblr in 
\citealt{kaspi2005}) to deduce $R_{\rm BLR}$ as comparison. To calculate the virial 
factors of the hidden BLRs in type 2 AGN sample, their extinction in narrow emission 
lines and the absorption in 2-10 keV X-ray luminosities should be also taken into 
account (see the references in Table \ref{tab:obj}). The extinction-corrected \oiii\ 
luminosities (corrected individually using the Balmer decrement of NLR 
for each object found in the literatures) and absorption-corrected hard 
X-ray luminosities compiled from the literatures, as well as their deduced virial 
factors (denoted as \foiii and \fx, respectively), are listed in Table \ref{tab:obj}.

For the uncertainties of virial factors, the dispersions of the empirical relationships 
we used are included. For \foiii, the primary uncertainties come from the 
dispersions of Equation \ref{eqn:loiii_tau} (0.31 dex, Figure \ref{fig:loiii_tau}). 
The uncertainties caused by the \oiii\ luminosity measurements are much smaller, and can be 
ignored. And we use the dispersion (62 percents) of the relation between \Lx and \rblr 
\citep{kaspi2005} to calculate the error bars of \fx. The uncertainties of \Lx 
mainly result from their intrinsic variations over time, and have been rolled  
into the dispersion of the \Lx -- \rblr relationship. Therefore, we do not include 
them explicitly.

\begin{deluxetable*}{lccccccccccccc}
\tabletypesize{\footnotesize}
\tablecaption{Virial factors\label{tab:obj}}
\tablewidth{\textwidth}
\tablehead{
	\colhead{Object} & \colhead{FWHM($\ha$)} & \colhead{Ref.} & & \colhead{$M_{\bullet}$} & 
	\colhead{Ref.} & & \colhead{$\log L_{\rm [OIII]}^a$} & \colhead{Ref.} & \colhead{\foiii} & & 
	\colhead{$\log \Lx$} & \colhead{Ref.} & \colhead{\fx} \\ 
	\cline{2-3} \cline{5-6} \cline{8-10} \cline{12-14} 
	& ${\rm (km\ s^{-1})}$ & & & $(M_{\odot})$ & & & $(\rm erg\ s^{-1})$ & & & & $\rm (erg\ s^{-1})$ & & 
}
\startdata
 Circinus  & $2300\pm 500$ & 1     & & $1.14^{+0.20}_{-0.20}\times10^6$ & 2 & & 41.17 & 3   & $ 0.17^{+ 0.15}_{- 0.15}$ & & 42.62 & 4 & $ 0.20^{+ 0.16}_{- 0.16}$ \\
 IC 2560   & $2100\pm 300$ & 1     & & $2.51^{+3.80}_{-1.51}\times10^6$ & 5 & & 40.19 & 6   & $ 1.83^{+ 3.11}_{- 1.79}$ & & 41.80 & 7 & $ 1.47^{+ 2.45}_{- 1.34}$ \\
 NGC 1068  & $4377\pm 300$ & 8     & & $8.39^{+0.44}_{-0.44}\times10^6$ & 9 & & 42.38 & 4   & $ 0.06^{+ 0.05}_{- 0.05}$ & & 43.02 & 4 & $ 0.25^{+ 0.16}_{- 0.16}$ \\
 NGC 2273  & $2399\pm 419$ & 10$^b$& & $8.61^{+0.91}_{-0.46}\times10^6$ & 11& & 41.13 & 4   & $ 1.28^{+ 1.02}_{- 1.02}$ & & 42.73 & 4 & $ 1.24^{+ 0.89}_{- 0.89}$ \\
 NGC 3393  & $5000\pm 600$ & 1     & & $1.57^{+0.98}_{-0.99}\times10^7$ & 12& & 42.04 & 3   & $ 0.15^{+ 0.14}_{- 0.14}$ & & 41.60 & 7 & $ 2.08^{+ 1.90}_{- 1.91}$ \\
 NGC 4388  & $4500\pm1400$ & 1     & & $7.31^{+0.17}_{-0.18}\times10^6$ & 11& & 41.85 & 4   & $ 0.11^{+ 0.10}_{- 0.10}$ & & 42.90 & 4 & $ 0.24^{+ 0.21}_{- 0.21}$  
\enddata
\tablecomments{
References: 1. \cite{ramos2016}, 2. \cite{greenhill2003}, \cite{kormendy2013}, 
3. \cite{bassani1999}, 4. \cite{marinucci2012}, 5. \cite{lasker2016}, 6. \cite{gu2006}, 
7. \cite{tilak2008}, 8. \cite{inglis1995}, 9. \cite{lodato2003}, \cite{kormendy2013}, 
10. \cite{moran2000}, 11. \cite{kuo2011}, \cite{kormendy2013}, 12. \cite{kondratko2008}, 
\cite{kormendy2013}. 
$^{a}$The Galactic and local extinction has been corrected. $^{b}$We measure the FWHM 
from the polarized spectrum in Ref. 10. We do not list the uncertainties of $L_{\rm [OIII]}$ 
and \Lx here, because we only use the dispersion of the empirical relationship to 
deduce the uncertainties of \foiii and \fx (see Section \ref{sec:data}).}
\end{deluxetable*}

\begin{figure*}
\includegraphics[width=\textwidth]{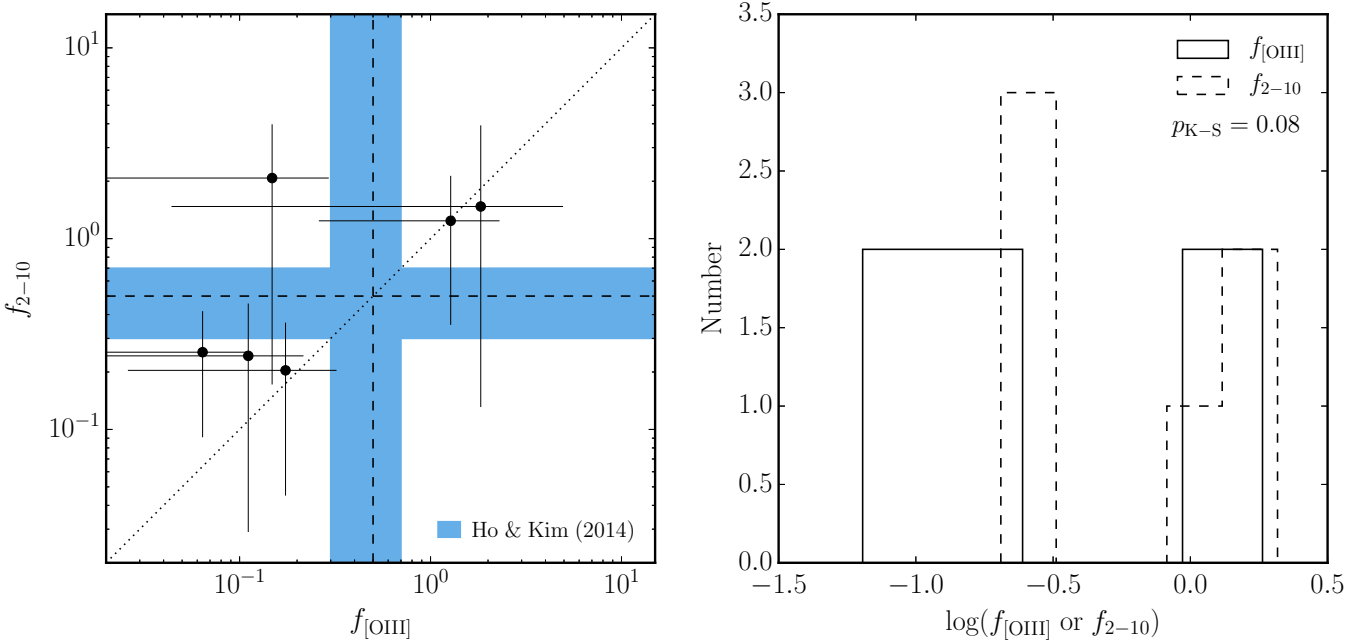}
\caption{Virial factors. The left panel is the comparison between 
\foiii and \fx. The dashed lines and blue regions are the virial 
factor and its uncertainty for AGNs with pseudobulges in 
\cite{ho2014}. The dotted line is the diagonal. The right panel 
shows the distributions of \foiii and \fx (solid lines for \foiii 
and dashed lines for \fx). $p_{\rm K-S}$ is the Kolmogorov -- Smirnof 
probability (see Section \ref{sec:data}).}
\label{fig:f_factor}
\end{figure*}

The comparison between the virial factors \foiii and \fx\ and their
distributions are provided in Figure \ref{fig:f_factor}. The mean values of
\foiii and \fx are 0.60 and 0.92, and their standard deviations are 0.69 and 0.73, 
respectively. The Kolmogorov -- Smirnof test shows the probability is 
8\% (It means, if the two distributions were identical, they would appear as 
discrepant as the observations indicate by chance in 8\% of the cases).
Considering that the size of the present sample is quite small and the scatter 
is moderately large, we think the resulting \foiii and \fx in the present sample 
are not significantly inconsistent with each other\footnote{In fact, in the present 
sample, only \foiii and \fx of NGC 3393 looks different. The relatively anomalous 
behavior of NGC 3393 could be explained by the very large variation in its X-ray 
luminosity (see Figure 1 in \citealt{fabbiano2011}).}. 
More observations are needed to clarify this point. The empirical relation between 
\Lx and \rblr we used to deduce \fx has more uncertainty \citep{kaspi2005}, 
we prefer to use \foiii in the following discussion.

We individually corrected the extinction of $L_{\rm [OIII]}$ for the six type 2 
AGNs (see the references in Table \ref{tab:obj}), but we adopt a characteristic 
E(B-V) value to establish the $R_{\rm BLR}$ - $L_{\rm [OIII]}$ relation for the 
RM AGNs. This characteristic correction method applies to a considerable size sample: 
(1) Adopting a characteristic value does not influence the slope of the 
$R_{\rm BLR}$ - $L_{\rm [OIII]}$ relationship, because the Balmer decrement 
of NLR does not correlate with \oiii\ luminosity \citep{shen2011}. 
(2) We have checked the effects of different characteristic extinctions, and 
find if we change it to 0.18 or 0.38, the \foiii\ only becomes 18\% smaller or 
22\% larger, respectively. These uncertainties are much smaller than the error bars. 
We searched $({\rm H\alpha/H\beta})_{\rm NLR}$ for the RM sample, and found the 
values for 20 objects which can be measured from their narrow lines reliably 
(see Table \ref{tab:hahb}). The median $({\rm H\alpha/H\beta})_{\rm NLR}$ for 
the 20 objects is 4.29, and the corresponding E(B-V) $=$ 0.31 which is similar 
to the value (0.28) used here. And also, 0.28 is consistent with the mean NLR 
extinction reported by \cite{heard2016} (see their Figure 3). Although we cannot 
find ${\rm (H\alpha/H\beta)_{\rm NLR}}$ reported for the entirety of the RM sample, 
there is no evidence that the average E(B-V) for the whole RM sample is significantly
different from the value we used here. Therefore, the characteristic extinction is 
reasonable for the present sample.

\begin{deluxetable}{lcc}
\tabletypesize{\footnotesize}
\tablecaption{Balmer Decrements\label{tab:hahb}}
\tablewidth{\textwidth}
\tablehead{
    \colhead{Object} & \colhead{$(\ha/\hb)_{\rm NLR}$} & \colhead{Ref.}
}
\startdata
SDSS J075101.42+291419.1 & 6.16 & 1 \\
Mrk 79                   & 2.68 & 2 
\enddata
\tablecomments{
References:
1. Obtained from the spectrum in SDSS, 
2. Obtained from the spectrum in NASA/IPAC Extragalactic Database.
(This table is available in its entirety in machine-readable form.)}
\end{deluxetable}

\section{Comparison with virial factor in type 1 AGNs}
\label{sec:compare}

In order to compare the obtained virial factors in the type 2 sample 
with the values in type 1 AGNs, we first provide some necessary background.

Once the time lag between continuum and emission line, as well as the emission 
line width, have been measured from RM, the mass of BH in an AGN can be determined 
using Equation \ref{eqn:mass} if the virial factor $f_{\rm vir}$ is known. Due to 
the fact that the velocity measurement could be either FWHM or $\sigma$, and the
velocity could be measured from mean spectra or rms spectra \citep{peterson1998},
there are four combinations of $f_{\rm vir}$ estimates ($f_{\rm vir}$ for FWHM in 
mean or rms spectra, or for $\sigma$ in mean or rms spectra). For simplicity, 
we designate them as \fmeanfwhm, \frmsfwhm, \fmeansigma and \frmssigma.

Assuming the velocity distribution of BLR is isotropic (which is likely not true), 
the simplest deduction from theory is that the virial factor for FWHM is 0.75 
\citep{netzer1990}. Observationally, the average virial factor for a sample can be
calibrated by comparing the RM objects with measurements of bulge stellar velocity
dispersion with the \msigma\ relation of inactive galaxies, if we assume 
that active and inactive galaxies follow the same \msigma\ relation. Based on a
sample of 14 AGNs with RM observations, \cite{onken2004} calibrated
$\frmsfwhm=1.4$ and $\frmssigma=5.5$ through the comparison with the \msigma
relation from \cite{ferrarese2000} and \cite{tremaine2002}. By adding the
observations from Lick AGN Monitoring Project (LAMP, e.g.,
\citealt{bentz2009}), \cite{woo2010} compiled a sample of 24 RM objects with
$\sigma_{*}$ measurements and found that $\log
\frmssigma=0.72^{+0.09}_{-0.10}$, if adopting the \msigma relation in
\cite{gultekin2009}, which is in good agreement with the value of
\cite{onken2004}. \cite{graham2011} improved the \msigma relation by including
more BH mass measurements of barred galaxies, and derived
$\frmssigma=3.8^{+0.7}_{-0.6}$, which is lower than the value in
\cite{woo2010} by a factor of 2, for their sample of 28 AGNs. When dividing 
the sample into barred and non-barred galaxies, they found \frmssigma would
be $2.3^{+0.6}_{-0.5}$ and $7.0^{+1.8}_{-1.4}$, respectively. Furthermore, they
accounted for the selection bias caused by the non-detection of
intermediate-mass BHs, and gave $\frmssigma=2.8^{+0.7}_{-0.5}$,
$2.3^{+0.9}_{-0.6}$ and $5.4^{+1.5}_{-1.2}$ for full, barred and non-barred
AGN samples. \cite{park2012} claimed that the discrepancy of 
virial factors in those previous works is mainly caused by the sample
selection effect, and preferred to use $\log\frmssigma=0.71\pm0.11$.
\cite{woo2013} investigated the assumption that active and
inactive galaxies follow the same \msigma relation, and demonstrated 
it is generally reasonable. \cite{grier2013} updated the RM 
sample, and added new $\sigma_*$ measurements of a few highly-luminous
quasars. They obtained $\frmssigma=4.31\pm1.05$ which is slightly lower than
Park et al.'s value but larger than that in \cite{graham2011}. Motivated by
the fact that galaxies with pseudoblges do not obey the \msigma relation of
classical bulges and ellipticals, \cite{ho2014} separated the calibration
by bulge types in galaxies and provided calibrated values of
[\fmeanfwhm, \fmeansigma, \frmsfwhm, \frmssigma]  = [$1.3\pm0.4$, $5.6\pm1.3$,
$1.5\pm0.4$, $6.3\pm1.5$] for classical bulges and ellipticals, and
[$0.5\pm0.2$, $1.9\pm0.7$, $0.7\pm0.2$, $3.2\pm0.7$] for pseudobulges.
More recently, \cite{woo2015} used the single-epoch spectra and 
the $R_{\rm BLR}$ -- $L_{5100}$ relationship to calibrate  
narrow-line Seyfert 1 galaxies, and found $\log\fmeanfwhm=0.05\pm0.12$ and 
$\log\fmeansigma=0.65\pm0.12$. 

Based on another technology, \cite{pancoast2014b} carried out
MCMC to reconstruct kinematics models of BLRs in five Seyfert 1 galaxies. 
It does not rely on \msigma\ relation, but uses RM data only and can provide 
$f_{\rm vir}$ estimates for individual objects. Adopting this an independent method, 
\cite{pancoast2014b} derives the averages $f_{\rm vir}$ 
for the five objects to be $\approx0.85$ and $\approx4.74$ corresponding 
for FWHM and $\sigma$, respectively. However, the values in individual objects 
are very different (by an order of magnitude). They also showed that 
$f_{\rm vir}$ strongly depends on the inclination angle of the BLR and is 
much larger in more face-on objects (from $\sim0.2$ for the viewing angle of 50 degrees 
to $\sim6$ for 10 degrees)\footnote{It should be noted that, besides the inclination 
angle, the thickness of BLR and some other dynamical parameters also influence 
the virial factor. Also \cite{pancoast2014b} reported that 4 of 5 objects show 
very thick BLRs, only the BLR of Mrk 1310 is relatively thin.}.

For the present Seyfert 2 sample, it is difficult to measure
the $\sigma$ of emission lines accurately in their polarized spectra given the
poor signal-to-noise ratio ($\sigma$ depends on accurate measurement of the wings 
of emission lines). However, FWHMs are more robustly measured. We
only estimate the virial factors based on FWHM. The average value of
$f_{\rm vir}$ in those Seyfert 2s is 0.60, and the dispersion is 0.69. Considering that
all of the objects in the sample are galaxies with pseudobulges
\citep{kormendy2013}, this number is in good agreement with the \fmeanfwhm and
\frmsfwhm for Seyfert 1 galaxies with pseudobulges in \cite{ho2014}.
Therefore, we can draw a preliminary conclusion, based on the 
present type 2 sample of limited size, that the virial factors in type 2 
AGNs are consistent with the values in type 1 AGNs of the same bulge type 
(pseudobulge).

Scattering regions in type 2 AGNs are mainly located along the polar axis 
(e.g., \citealt{antonucci1983, capetti1995}), and provide a new perspective 
to BLRs from pole-on direction (small viewing angles). The similar value of virial 
factors, which are found from the polarized spectra of type 2 AGNs, as in type 1 
objects indicate that the geometry and kinematics of BLR are similar in type 1 
and type 2 AGNs of the same bulge type (pseudobulge). And the small virial 
factors ($\foiii=0.60$) found from the polarized spectra demonstrate that the 
geometry of BLRs in the present sample tend to be very thick disks (thicker 
than the results in \citealt{pancoast2014b} because of our smaller $f_{\rm vir}$) or 
are perhaps even isotropic to some extent. 

It should be noted that thermal motion of the free electrons in scattering 
regions could provide an additional broadening to the polarized emission lines, 
and further, influence the estimates of virial factors to some extent. 
\cite{miller1991} shows that, consistent with the conclusion of \cite{antonucci1985}, 
thermal electrons dominate the scattering process in NGC 1068 and broaden the polarized 
emission lines. Unfortunately, similar information is unavailable in 
other AGNs. On the other hand, lower temperature dust grains could also 
contribute to the scattering, and may dominate at least in some objects. 
In such case, the additional broadening from the scattering particles is not an issue. 
The virial factor of NGC 1068 is low compared to the other 5 objects (Table \ref{tab:obj}). 
If the width from the dust scattering were used, it would be more similar. The detailed 
nature of scattering and the corresponding broadening still remain open questions and 
need to be investigated in the future.

\section{Summary}
\label{sec:summary}

In this work we compile a sample of six Seyfert 2 galaxies with 
spectropolarimetric observations and dynamical BH mass measurements, and 
derive their virial factors from the FWHMs of the polarized BELs, in order to investigate the 
kinematics and geometry of hidden BLRs in type 2 AGNs. Generally, the virial factors 
estimated in the different ways (from the luminosities of \oiii\ and X-ray) are in 
agreement. The average of the derived 
virial factors is 0.60 in the present sample, which is similar to the value of 
type 1 objects with pseudobulges \citep{ho2014}. It implies that (1) the geometry 
and kinematics 
of BLR are similar in type 1 and type 2 AGNs of the same bulge type (pseudobulge) 
and (2) the geometry of BLRs in type 2 AGNs tend to be a very thick disk or isotropic. 
In the future, more spectropolarimetric observations would make it possible to 
explore the properties of hidden BLRs in more details, and further, to investigate 
the dependency of hidden BLR kinematics and geometry on black hole masses or accretion 
rates of AGNs.

\acknowledgements{ 
We acknowledge the support of the staff of the Lijiang 2.4m telescope. 
Funding for the telescope has been provided by CAS and the People's Government of Yunnan 
Province. This research is supported  
by NSFC grants NSFC-11503026, -11173023, and 
-11233003, and a NSFC-CAS joint key grant U1431228,  
by the CAS Key Research Program through KJZD-EW-M06, by National Key Program 
for Science and Technology Research and Development (grant 2016YFA0400701), and by 
Key Research Program of Frontier Sciences, CAS, Grant NO. QYZDJ-SSW-SLH007.
}

\end{document}